\documentclass[floatfix,aps,prd,twocolumn,letterpaper,lengthcheck,superscriptaddress,showpacs,amssymb,amsmath,amsfonts,nofootinbib,altaffilletter,nopreprintnumbers,showpacs,longbibliography]{revtex4-1}
\usepackage[utf8]{inputenc}
\usepackage{xspace}
\usepackage{acronym}
\usepackage{booktabs}
\usepackage{todonotes}
\usepackage{appendix}
\usepackage{graphicx}
\usepackage{amssymb}
\usepackage{color}
\usepackage{ulem}
\usepackage{wasysym}
\usepackage{url}
\usepackage[colorlinks, linkcolor=blue, pdfborder={0 0 0}, breaklinks=true]{hyperref}
\definecolor{lightgray}{gray}{0.9} % table alternating line colors
\usepackage{tabularx}
\usepackage{multirow}
\usepackage{scrextend}
\usepackage{textgreek}
\usepackage{graphicx}
\usepackage[caption=false]{subfig}
\usepackage{mathtools}
\usepackage{float}
\usepackage{soul}
\usepackage{booktabs}
\usepackage{enumitem}
\usepackage{dcolumn}% Align table columns on decimal point
\usepackage{bm}% bold math

\usepackage{lineno} % line numbers
\newcommand{\qm}[1]{\textcolor{black}{#1}}

\DeclarePairedDelimiterX{\norm}[1]{\lVert}{\rVert}{#1}
\begin{document}

\preprint{APS/123-QED}
%%%%----------------------------------------------------
\title{Fast Waveform Generation for Gravitational Waves using Evolutionary Algorithms} 
%%%%----------------------------------------------------

\author{Quirijn Meijer}\email[Corresponding author: ]{r.h.a.j.meijer@uu.nl}
\affiliation{Institute for Gravitational and Subatomic Physics (GRASP), Department of Physics, Utrecht University, Princetonplein 1, 3584CC Utrecht, Netherlands}
\affiliation{Nikhef, Science Park 105, 1098XG Amsterdam, The Netherlands}

%\author{Harsh Narola}
%\affiliation{Institute for Gravitational and Subatomic Physics (GRASP), Department of Physics, Utrecht University, Princetonplein 1, 3584CC Utrecht, Netherlands}
%\affiliation{Nikhef, Science Park 105, 1098XG Amsterdam, The Netherlands}

\author{Sarah Caudill}
\affiliation{Department of Physics, University of Massachusetts, Dartmouth, Massachusetts 02747, USA}
\affiliation{Center for Scientific Computing and Data Science Research, University of Massachusetts, Dartmouth, Massachusetts 02747, USA}

\date{\today}

%%%%---------------------------------------------------
\begin{abstract}
\noindent Gravitational-wave analyses depend heavily on waveforms that model the evolution of compact binary coalescences as seen by observing detectors. In many cases these waveforms are given by waveform approximants, \qm{models that approximate the amplitude and phase of the waveform at a set of frequencies}. Because of their omnipresence, improving the speed at which approximants can generate waveforms is crucial to accelerating the overall analysis of gravitational-wave detections. An optimisation algorithm is proposed that can select at which frequencies in the spectrum an approximant should compute the power of a waveform, and at which frequencies the power can be safely interpolated at a minor loss in accuracy. The algorithm used is an evolutionary algorithm modeled after the principle of natural selection, iterating frequency arrays that perform better at every iteration. As an application, the candidates proposed by the algorithm are used to reconstruct signal-to-noise ratios. It is shown that the IMRPhenomXPHM approximant can be sped up by at least $30\%$ at a loss of at most $2.87\%$ on the drawn samples, measured by the accuracy of the reconstruction of signal-to-noise ratios. The behaviour of the algorithm as well as lower bounds on both speedup and error are explored, leading to a proposed proof of concept candidate that obtains a speedup of $46\%$ with a maximum error of $0.5\%$ on a sample of the parameter space used.
\end{abstract}

%%%%---------------------------------------------------
\maketitle

%%%%---------------------------------------------------
\section{Introduction}\label{sec:intro}

Gravitational waves were first predicted well over a hundred years ago by Heaviside, Poincar\'{e} and Einstein, with Einstein famously remaining skeptical of their existence throughout his life. Einstein believed that the amplitude of gravitational waves reaching earth would be so small that their detection could never be within human reach. Despite the accurate prediction of their magnitude, the LIGO-Virgo collaboration succeeded in making their first detection of a gravitational wave in 2015 \cite{gw150914}. Since, the LIGO-Virgo-KAGRA collaboration have confirmed detection of over $90$ gravitational waves \cite{PhysRevX.9.031040, PhysRevX.11.021053, theligoscientificcollaboration2021gwtc3, theligoscientificcollaboration2022gwtc21}. Gravitational waves are now used to study black holes, neutron stars and other cosmological sources of gravitational waves.

A prominent routine in the analysis of gravitational wave data is matched filtering \cite{helstrom1960statistical}, a process in which templates that predict the time or frequency evolution of a gravitational wave are convolved with incoming data \cite{helstrom1960statistical}. In order to cover many possible source systems, templates for different systems are collected in template banks \cite{PhysRevD.44.3819}. In turn, the templates populating this bank are generated by waveform approximants. Waveform approximants \cite{Buonanno_1999, Pan_2014, Estell_s_2022, PhysRevD.93.044006, PhysRevD.93.044007} serve to approximate the waveform of a system either in the time or frequency domain and are faster alternatives to the more precise waveform simulations from numerical relativity \cite{PhysRevD.104.084038, Aasi_2014, Jani_2016, Healy_2017, Healy_2022}. Approximants balance the need for both speed and accuracy, and therefore approximate a waveform to a reasonable accuracy \cite{Babak_2017} within a reasonable time. Other applications for approximants are in parameter estimation \cite{RevModPhys.94.025001}, where waveforms are generated during runtime, so that there too computational speed is crucial. If waveforms are computed during runtime this is called online computation, whereas it is called offline computation if this task is delegated to the pre-processing stage.

Since the Fourier transform provides a duality between the time domain and frequency domain, the domain of analysis is a choice. Sometimes it is preferential to work in the frequency domain, where a different perspective on data is allowed, or computations are easier to perform. This is the case for the methods presented in this paper, where the frequency contents can be considered the building blocks of a waveform. A waveform approximant generating a waveform in the frequency domain will do so by solving sets of equations at individual frequencies in the spectrum \cite{PhysRevD.103.104056}. Doing so is costly, notably when waveforms are generated online during the runtime of a procedure, but also in generally large undertakings such as template bank validation where percentual changes make a large absolute difference \cite{PhysRevD.80.104014, PhysRevD.89.084041, PhysRevD.94.024012, mcisaac2023search}. One way of reducing the cost of computation would be to evaluate a waveform on as small an array of frequencies as possible, resorting to alternative, faster procedures to determine the behaviour at the remaining frequencies. This is the premise of this paper.

With this line of reasoning, a method that can speed up waveform approximants is calling the solver for a subset of the frequencies, and interpolating the values of the frequencies in the complement within the frequency spectrum. Finding where in the spectrum to solve and where to interpolate is then an optimisation problem, in particular one with a very large solution space that has size growing exponentially in the number of frequencies. One way to efficiently find a (local) optimum for this problem is through the use of evolutionary algorithms \cite{7955308}.

Evolutionary algorithms mimic natural evolution, where successive generations are expected to be better adapted to their environments. This principle translates to an algorithm that at each iteration tests a set of solutions, collectively called a generation, and then forms a new generation that contains better solutions to the problem. Evolutionary algorithms, sometimes also referred to as genetic algorithms, have been applied to gravitational-wave data analysis in other work \cite{Cavaglia_2020, CiCP-25-963, skliris2022realtime}.

Alternative methods with the same motive have been proposed, such as multi-band template interpolation \cite{Vinciguerra:2017ngf} and adaptive frequency resolution \cite{Morisaki:2021ngj}. However, both methods make assumptions on the frequency evolution of the waveform to reduce the number of frequencies at which the power is computed, and adaptive frequency resolution in particular was shown to be accurate mostly for relatively loud signals. Evolutionary algorithms require no assumptions to be made and are therefore more widely applicable, with no specific boundary on the loudness. Other related methods include the work done on reduced order quadratures \cite{PhysRevD.94.044031, PhysRevLett.114.071104, PhysRevD.108.123040} and meshfree approximations \cite{pathak2024fast}.

The application of evolutionary algorithms to the optimisation problem of waveform generation is studied in this paper as a proof of concept, proposing an algorithm, and studying both the time speed-up and the coverage of the parameter space. Coverage here means the generalisibility of the frequency array output by the algorithm to samples taken from the parameter space, or on what subset of the parameter space the interpolation from the output frequency array gives accurate approximations.

The contents of this paper are ordered as follows. In Sec. \ref{sec:approx} approximants are briefly introduced. In Sec. \ref{sec:evoalgo} evolutionary algorithms are treated. Sec. \ref{sec:methodology} then outlines the used methodology, including the design of the algorithm and the test cases for the output. Results are reported in Sec. \ref{sec:results} before the work is reviewed in Sec. \ref{sec:conclusions}.

%%%%---------------------------------------------------
\section{Approximants}\label{sec:approx}

In a compact binary coalescence, as two objects such as black holes or neutron stars orbit each other, energy is lost through gravitational wave emission \cite{10.1093/acprof:oso/9780198570745.001.0001}. A typical process consists of the inspiral, during which the objects inspiral closer together, the merger, at which point the objects coalesce, and the ringdown, where the single resultant object settles down into its nature \cite{10.1093/acprof:oso/9780198570745.001.0001}. 

In the case of a non-eccentric binary black hole system, its evolution is uniquely defined by two mass parameters, six spin parameters, two sky location parameters along with the phase of coalescence, inclination, polarisation angle, luminosity distance and the time of arrival \cite{PhysRevD.49.6274}. These parameters are collected in a $15$-dimensional parameter space denoted as $\mathcal{P}$. Since there are no a priori restrictions on the spin parameters, a binary black hole can be defined by a set of parameters with arbitrary spin orientations. If this is the case, the total angular momentum and orbital angular momentum may be misaligned, causing precession of the orbital plane \cite{PhysRevD.49.6274}. It is then said that this system demonstrates precession, appearing in the waveform through amplitude and phase modulations \cite{PhysRevD.49.6274, PhysRevD.86.104063}. 

If instead of a binary black hole a system containing a neutron star is considered, an additional parameter describing the tidal deformability of the star is added to this space \cite{PhysRevD.81.123016}. This paper will however restrict itself, and therefore the parameter space, to the case of black holes.

A waveform approximant defined on these parameters can then be interpreted as a function:

\begin{equation}
    A_{\mathbb{F}}: \mathcal{P} \rightarrow D(\mathbb{F}, \mathbb{C})
\end{equation}

\noindent that maps a given set of parameters into the set of discrete complex functions on the frequency spectrum $\mathbb{F} \subset \mathbb{R}$, which should not be confused with an algebraic field. Note that $D(\mathbb{F}, \mathbb{C})$ can alternatively be written as $\mathbb{C}^{\mathbb{F}}$, and that the notation $A_{\mathbb{F}}$ specifically includes the spectrum $\mathbb{F}$. Such a waveform approximant is typically the best possible option for the modeling of a system's evolution short of an analytical or numerical solution of the Einstein field equations with stress-energy-momentum tensor defined by the parameters of choice. This will be discussed in more detail shortly.

Waveforms approximants can include what are known as higher order modes \cite{10.1093/acprof:oso/9780198570745.001.0001}. In the time domain, a waveform $h(t)$ with polarisations $h_{+}(t)$ and $h_{\times}(t)$ can be written in multipole expansion as a linear combination of the spherical harmonics $Y_{l, m}^{-2}$ of weight $2$ \cite{atkinson2012spherical, kosmann2009groups, PhysRevD.97.023004, Pratten:2020ceb, Hannam:2013oca, Garcia-Quiros:2020qpx, PhysRevD.79.104023, 10.1093/acprof:oso/9780198570745.001.0001}:

\begin{equation}
	h_{+}(t) + ih_{\times}(t) = \sum_{l \leq 2} \sum_{m = -l}^{m = l} Y_{l, m}^{-2}(\iota, \varphi) h_{l, m}(t),
\end{equation}

\noindent where the $h_{l, m}(t)$ are the modes of gravitational-wave emission. Note that the spherical harmonics depend on the polar angle ($\iota$) and the azimuthal angle ($\varphi$), that partially define a frame of reference for the binary black hole with origin in the center of mass of the system. In typical searches it is assumed the quadrupole $(l, m) = (2, \pm 2)$ mode dominates the expansion \cite{PhysRevD.97.023004}, and moreover, the maxima of $Y_{2, \pm2}^{-2}(\iota, \varphi)$ at $\iota = 0, \pi$ coincide with face-on sources for which the dominant direction of gravitational-wave emission is aligned with the normal vector to the detector plane. For these reasons the quadrupole mode may serve as a good approximation for many waveforms, akin to a lower-order Taylor expansion. However, for other systems the higher order modes may contribute significantly to the expansion \cite{PhysRevD.97.023004}, for instance for edge-on systems where $\iota = \pi/2$. This value of $\iota$ corresponds to a minimum of $Y_{2, \pm2}^{-2}(\iota, \varphi)$, but a maximum of many other spherical harmonics. Approximants that take into account modes for $l, \left| m \right| > 2$ are said to include higher order modes.

The most accurate way of obtaining waveforms is through numerical relativity \cite{baumgarte2010numerical, PhysRevLett.95.121101, PhysRevD.104.084038, Aasi_2014, Jani_2016, Healy_2017, Healy_2022}, where the Einstein field equations are solved numerically. Although precise, these methods are very computationally expensive. Approximants arise through the need to obtain waveforms that approximate these solutions at a reasonable accuracy, but with far greater speed. As a calibration benchmark therefore, numerical relativity remains an important part of the study of approximants. In practice, different families of waveform approximants exist, such as the Taylor family \cite{PhysRevD.80.084043, PhysRevD.99.124051} for post-Newtonian approximations, the EOB (effective-one-body) family \cite{Buonanno:1998gg, PhysRevD.106.024020} that reduces two-body dynamics to the case of a corresponding single body, and the phenomenological family \cite{Garcia-Quiros:2020qpx, PhysRevD.102.064001, PhysRevD.103.104056}, with the latter being the fastest as they are designed specifically to be faster than the EOB family while retaining similar accuracy. A sophisticated example from the phenomenological family is the frequency-domain approximant IMRPhenomXPHM \cite{PhysRevD.103.104056}, that can compute the waveforms of precessing system with the inclusion of higher order modes. Internally, this approximant fits behaviour to individual frequencies before constructing a precessing waveform through modulation by the twisting up procedure \cite{PhysRevLett.113.151101}. In this paper IMRPhenomXPHM is used as the approximant of choice, as it is fast, commonly used online, and incorporates both precession and higher order modes. Note however that the methods presented in this paper are agnostic to this choice and are applicable to all approximants.

\begin{figure*}[!]
    \includegraphics[width=1\textwidth]{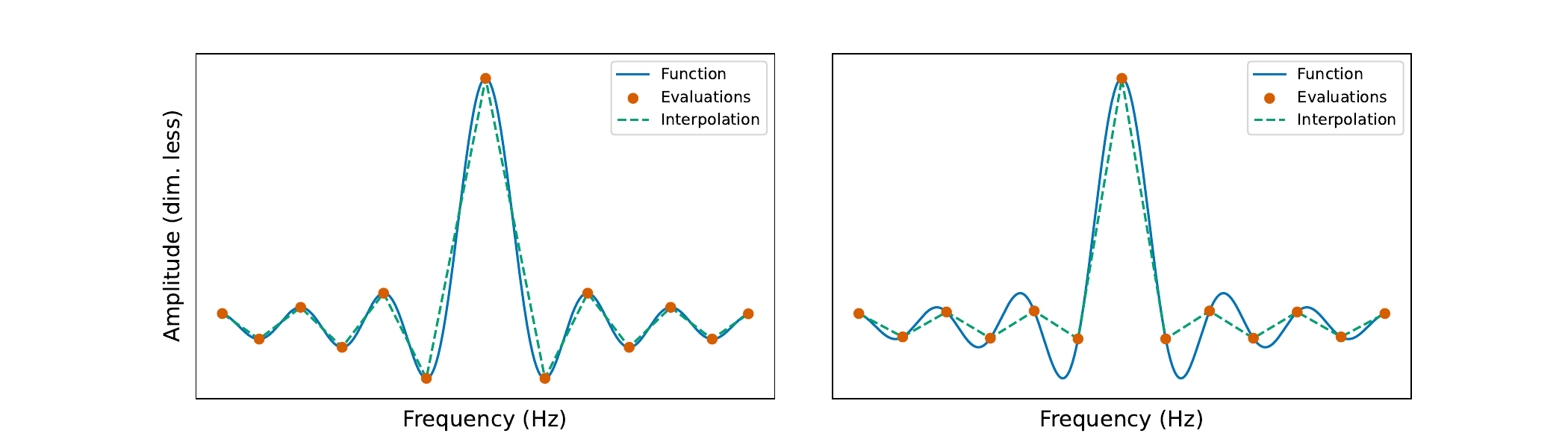}%
    \caption{A damped sinusoid interpolated on two different frequency arrays. The left frequency array consists of the turning points of the function, whereas the right frequency array is uniformly spaced. This figure shows both how frequency arrays can be realised differently, as well as the importance of the choice of array. Although both contain the same number of frequencies, corresponding to the evaluations in the plots, the left interpolation is clearly a better approximation to the true function.}
    \label{fig:frequency_array_example}
\end{figure*}

As will be demonstrated in this paper using IMRPhenomXPHM for $A_{\mathbb{F}}$, obtaining $A_{F}$ for a subset $F \subset \mathbb{F}$ and interpolating $A_{F}$ to $A_{F}^{*}$ on $\mathbb{F}$, where the asterisk indicates interpolation, is faster on a per-frequency basis than solving for all frequencies in $\mathbb{F}$, yet still accurate. The subset $F$ will be referred to as the frequency array, and two coarse examples of such frequency arrays for a damped sinusoid are shown in Fig. \ref{fig:frequency_array_example}. The convention of asterisks indicating linear interpolations will be upheld. Notation will however be shortened, with the exact waveform $A_{\mathbb{F}}(\theta)$ being denoted by $h(\theta)$, and the approximate waveform $A_{F}^{*}(\theta)$ by $h_{F}^{*}(\theta)$. If the choices of $\theta$ and $F$ are implicit, the notation can be further reduced to $h$ and $h^{*}$ respectively.

Since this speed-up is obtained through interpolation and not algorithmically, $h$ and $h^{*}$ can not be expected to be equal, and a loss in accuracy will be incurred. Two metrics will be used to measure this loss. The first is the residual sum of squares \qm{(RSS)} for the real and imaginary parts of $h$ and $h^{*}$. This metric is used by the evolutionary algorithm internally. The second metric is based on a widely used quantity, the signal-to-noise ratio (SNR).

The concept of SNR comes from signal processing, and is defined for a given power spectral density (PSD) as the normalisation with respect to $u$ of the inner product:

\begin{equation}
    \langle u, v \rangle = \frac{4}{T} \; \sum_{f \in \mathbb{F}} \frac{u(f)\overline{v(f)}}{\textup{PSD}(f)},
    \label{eq:innerproduct}
\end{equation}

\noindent where $T$ is the duration of the waveform, the bar denotes the complex conjugate, and for the applications in this paper, the PSD describes the sensitivity of the detector at different frequencies. The value of $\langle u, v \rangle$ can be interpreted as a quantisation of the presence of $u$ in $v$, as a ratio of signal to noise.

%%%%---------------------------------------------------
\section{Evolutionary Algorithms}\label{sec:evoalgo}

Evolutionary algorithms \cite{7955308} are inspired by biological evolution, where natural selection leads populations to evolve to adapt to their environment. As such, this class of algorithms is used as approximation schemes for difficult optimisation problems. As in genetics, data is represented internally by genotypes that contain the hereditary information of an organism, whereas the phenotype refers to the externally observable characteristics. In practice this means a solution is encoded in a genotype, with the full solution phenotype being constructed from the genotype. Besides an injective mapping from genotypes and phenotypes, an evolutionary algorithm is governed by the following set of functions.

\begin{enumerate}
    \item An objective function or fitness evaluation function $O: G \rightarrow \mathbb{R}$ that assigns a genotype a fitness score measuring performance;
    \item A (stochastic) parent selection function $P$ that draws two parents from the set of genotypes and;
    \item A recombination function $R: G \times G \rightarrow G$ that combines two genotypes to form a new genotype. 
\end{enumerate}

\noindent Note that the recombination function can include stochastic mutations of the genotype, useful to add small perturbations that might nudge a solution from a non-optimal potential well in the objective field. The steps in an evolutionary algorithm are then summarised as follows.

\begin{enumerate}
    \item A first generation of solutions is instantiated;
    \item Using $O$, each genotype is assigned a fitness;
    \item A new generation is populated through repeated calls to the composition function $R \circ P$;
    \item The above steps are repeated until a set of stopping conditions has been met.
\end{enumerate}

\noindent Examples of stopping conditions include reaching a pre-set number of generations, or the stalling of variation and improvement within the generations.

In general, evolutionary algorithms offer a method of exploring a solution space where the solutions can be complicated objects. Under the right assumptions, each subsequent generation of solutions is expected to improve on the previous generation, leading to well-performing solutions in practice, even if the determination of a global optimum is not guaranteed. A famous example is the evolved antenna by NASA, designed in automation for the detection of unusual radiation patterns \cite{nasaantenna}. 

Few theoretical results that explain the effectivity and convergence of evolutionary algorithms are known. However, a notable theorem often cited is the Schema theorem \cite{ALTENBERG199523} that states better adapted solutions are likely to drive up the average fitness of successive populations.

%%%%---------------------------------------------------
\section{Methodology}\label{sec:methodology}

With the optimisation problem and solution method defined, this section discusses the methodology, split into three subsections. The pre-processing section discusses all considerations that preceed the introduction of the evolutionary algorithm to the problem. Following this, the hyperparameters that define the evolutionary algorithm for the established problem of finding the optimal frequency arrays are discussed in the second subsection, with the third subsection offering the performance measures used to evaluate the proposed solutions. Three such solutions will be considered.

In typical applications the frequency spectrum $\mathbb{F}$ is a uniformly spaced set, and in the tests presented here the set is defined by upper and lower limits of $0$ and $2048$ Hz respectively, with a sampling frequency of $4$ kHz and a duration of $16$ seconds.

\subsection{Pre-processing}\label{subsec:preprocessing}

In order to obtain a faster rate of convergence for the evolutionary algorithm, reference events are used to populate initial frequency arrays during pre-processing. An event is defined here as a fixed set of parameters, and in the case of a catalogued candidate, the parameters used are retrieved from the parameter estimation data releases \cite{PhysRevX.9.031040, PhysRevX.11.021053, theligoscientificcollaboration2022gwtc21, theligoscientificcollaboration2021gwtc3}. Three such reference events are tested. First, GW150914 \cite{gw150914}, as it was a landmark detection without additional complexities. Second and third are GW190412 \cite{PhysRevD.102.043015} and GW190814 \cite{Abbott_2020} since there is evidence for the contribution of higher order modes in these events, as well as precession \cite{PhysRevResearch.2.043096, PhysRevD.106.023019}. For this proof of concept, only the plus-polarised components are used.

Starting from the parameters of a reference event, the parameters other than the chirp mass are fixed, and the chirp mass is varied by $\pm 5$ solar masses to generate a number of different waveforms on $\mathbb{F}$. The choice to vary the chirp mass was made as it induces a large variation in the phase of the resultant waveforms \cite{PhysRevD.49.2658}. Each of the waveforms are normalised in amplitude to $[-1, 1]$, and the turning points in the frequency domain are added to a set $B$ called the background or prior. Every frequency array for this event will include this ever-present background of frequencies. This is done under the assumption that the evolutionary algorithm finding and including turning points is inevitable, since a piecewise linear interpolation between the turning points of a function generally gives a very well performing approximation to the full function. The inclusion of this prior information in pre-processing greatly reduces the work the evolutionary algorithm would have to perform in identifying these frequencies during a random walk. Choosing the right number of waveforms, or equivalently the permitted cardinality of $B$, is an example of bias-variance tradeoff \cite{10.5555/1162264}. For the tests described here, $5$ waveforms were used, including the reference event.

Finally, all frequencies under $20$ Hz were removed as candidates for inclusion in a genotype. The result of these pre-processing steps is one set of frequencies (the background) per event that will be present in all frequency arrays learned on the corresponding event, as well as one permissible set of frequencies to include in the genotypes for each. 

\subsection{Hyperparameters}\label{subsec:hyperparameters}

Given the pre-processed data, the hyperparameters uniquely define the evolutionary algorithm up to stochasticity. In this subsection the different hyperparameters are treated, guided by the context wherein they arise. The reader is recommended to look ahead at Table \ref{table:hyperparameters} before reading this subsection in order to get an overview of how this subsection is structured.

\begin{figure}[!]
    \includegraphics[width=1\columnwidth]{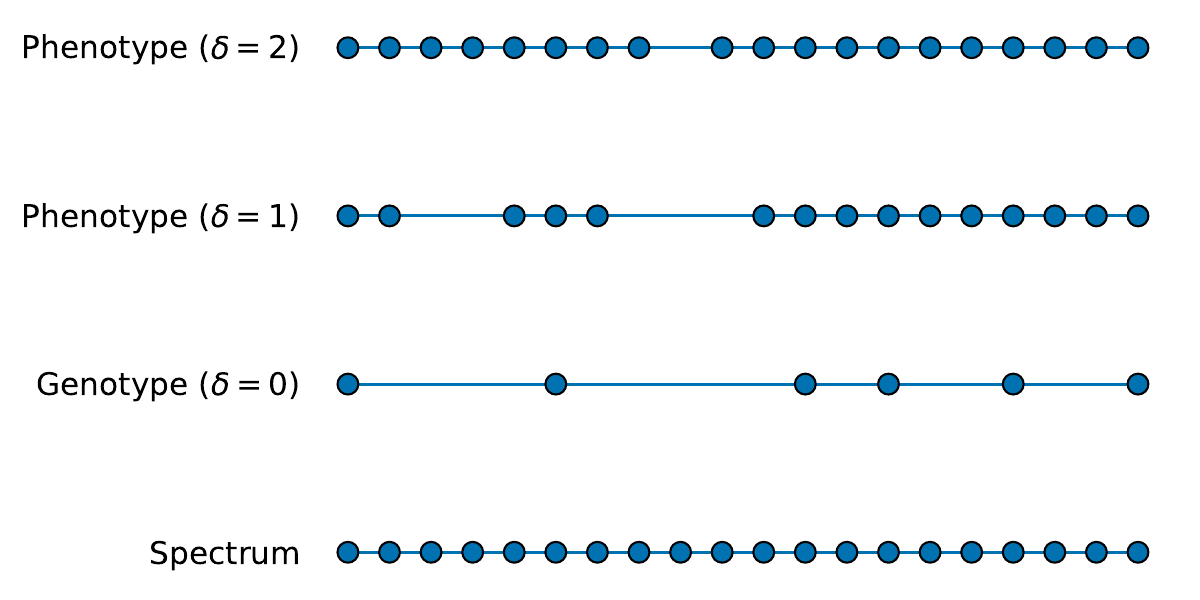}%
    \caption{An example genotype and corresponding phenotypes for a spectrum of subsequent natural numbers with empty background. For $\delta = 0$ the genotype and phenotype coincide. With $\delta = 1$ nearby frequencies are added to the genotype to form the phenotype, and with $\delta = 2$ additional nearby frequencies are added. This figure shows the hyperparameter $\delta$ controls the confidence with which the algorithm includes frequencies in the solution phenotypes.}
    \label{fig:phenotypes}
\end{figure}

Recall that for a set $S$ and a subset $T \subset S$, an indicator vector $I_{T}$ is a binary vector indexed by the elements of $S$ such that $I_{T}(s) = 1$ if and only if $s \in T$. Genotypes, as subsets of the frequency spectrum, are encoded in indicator vectors over $\mathbb{F}$, and a phenotype is constructed from such a genotype by adding the indicated frequencies to the frequencies in the background $B$. This map therefore realises a frequency array as a phenotype. Additionally, a parameter $\delta \in \mathbb{R}_{\geq 0}$ representing the confidence of the algorithm can be set: if a frequency $f$ is included in the genotype, then the elements of the set:

\begin{equation}
    \{ f' \in \mathbb{F} : \left| f - f' \right| \leq \delta \}
\end{equation}

\noindent will be included in the phenotype. Note that this means that with $\delta = 0$ only the frequency $f$ itself is included in the phenotype, therefore reducing to the special case described above. The phenotype (or frequency array) construction process for different values of $\delta$, with $B$ set to the empty set for simplicity, is shown in Fig. \ref{fig:phenotypes}.

For a genotype $G$ (with corresponding phenotype $F$) and a reference event waveform $h$, the objective function is defined as:

\begin{equation}
    O(G) = \sum_{f \in \mathbb{F}-F} \left| h(f) - h^{*}_{F}(f) \right| + w \cdot \# F,
    \label{eq:objective}
\end{equation}

\noindent where $\left| \cdot \right|$ denotes the modulus and the hash denotes set cardinality. The sum of moduli amounts to the residual sum of squares of the exact and approximate waveforms for both the real and complex parts, as can be seen from expanding the modulus for a single frequency. This sum can therefore also be expressed as:

\begin{equation}
    \textup{RSS}(\textup{Re}[h], \textup{Re}[h^{*}_{F}]) + \textup{RSS}(\textup{Im}[h], \textup{Im}[h^{*}_{F}]).
\end{equation}

\noindent An additive penalty, weighed by a variable $w$, is added for the phenotype size. One should take note of the range for the summation in the objective function defined in Eq. \ref{eq:objective}: for all frequencies in $F$ the exact waveform coincides with the approximate waveform, meaning these terms reduce to zero, and therefore do not have to be computed.

The complexity of the solution space, and therefore the need for optimisation methods like evolutionary algorithms, is demonstrated through the objective function. A first approach to optimisation could be by way of a greedy algorithm \cite{cormen01introduction}, where at every iteration the frequency $f'$ corresponding to the largest residual, or equivalently the largest term in the sum, is added to the phenotype. However, due to the addition of this frequency, the approximate waveform is changed to

\begin{equation}
    h^{*}_{F \cup \{ f' \}},
\end{equation}

\noindent and the terms of the sum in the objective function will have to be recomputed. At the same time the cross-section of the (very high-dimensional) solution space containing solutions without the frequency $f'$ is pruned. There is no guarantee that the frequency $f'$ that is now included is part of the optimal solution as this objective function has no optimal substructure \cite{cormen01introduction}, and no choice of frequency is ever reconsidered. It follows that the greedy algorithm would now be barred from finding a global optimum, whereas an evolutionary algorithm could rectify this misstep.

If a phenotype frequency array is used for the computation of the SNR, the error and rate of convergence are quantifiable, and minimising the sum of moduli in the objective function is equivalent to minimisation of the error term. Assume $D$ is an arbitrary but fixed strain of data, and $h$ is a waveform. It then follows from the linearity of the inner product that the error $E(h)$ for this waveform is given by:

\begin{equation}
    E(h) = \langle h, D \rangle - \langle h^{*}, D \rangle = \langle h - h^{*}, D \rangle.
    \label{eq:error}
\end{equation}

\noindent Since taking the limit of an inner product is interchangeable with taking the limit of an argument, and the continuous inner product will tend to $0$ if either argument tends to $0$, it is given that $E(h) \rightarrow 0$ if $h - h^{*} \rightarrow 0$. The latter is exactly the case if $\left| h - h^{*}_{F} \right| \rightarrow 0$ in the $\textup{L}^{1}$-norm.

Once the objective function has been used to evaluate the fitness of the genotypes in a generation, a new generation of offspring can be formed. The parent selection function $P$ is implemented according to fitness proportionate selection with elitism \cite{6791924, 10.1093/oso/9780195099713.001.0001}. Elitism refers to the carrying over of the top percentage of genotypes (by fitness ranking) to the next generation, and fitness proportionate selection means that the higher a genotype is ranked, the higher the probability will be that it is drawn as a parent by $P$. The probability measure so constructed on the genotypes sorted by ascending fitness is given by $\mathbb{P}(i) = i \cdot p$ with

\begin{equation}
    p = \frac{1}{\sum_{j = 1}^{g} j} = \frac{2}{g (g + 1)}
    \label{eq:probability}
\end{equation}

\noindent for the candidate ranked at position $i$, and $g$ the fixed population size of a generation. The use of this measure ensures that the lowest ranked genotype (at $i = 1$) has the lowest probability of being chosen as a parent, with every higher ranked genotype having a proportionally higher probability of being selected. The second equality in Eq. \ref{eq:probability} follows by substituting in Gauss' summation formula. Note too the importance of the generation size, as for larger values the probabilities will be  more evenly distributed over the candidates.

Finally, the recombination function $R$ is defined as the sum of the two genotypes, assuming the genotypes to be a boolean algebra. This means that a frequency will be present in the offspring if and only if that frequency is present in at least one of the parent genotypes.

Due to a symmetry argument applied to the phenotype construction process, the algorithm can be ran both forwards and backwards. In the forwards case, the phenotype grows in size towards the full frequency spectrum as frequencies are added in each iteration of the algorithm. In the backwards case the frequency spectrum is reduced in size towards the background by removing frequencies in each iteration. As a rule of thumb the forwards case is better suited to finding smaller frequency arrays, whereas the backwards case is better suited to finding frequency arrays that result in a low error. \qm{This is a result of the stopping conditions typically being met before the frequency array reaches half the size of the frequency spectrum.} Note that the parameter that dictates the direction stands in direct relation to the penalty $w$ included in Eq. \ref{eq:objective}. In both cases the algorithm needs to be incentivised to move through the solution space by the penalty. In for instance the backwards direction, if $w$ were to be set to $0$, there would be no reason to remove any frequency from the phenotype, as doing so would only lead to a loss in accuracy. Setting $w > 0$, the removal of this same frequency might result in an increased fitness, as long as the loss in accuracy is less than the penalty $w$ for including the frequency.

The evolutionary algorithm that has now been established can only be run once an initial generation is given. Initialisation of this first generation is done at random through a uniform distribution on the set of allowed frequencies. A generation of genotypes is formed by drawing a sample of frequencies of size $s$ from this distribution, repeating this until $g$ such genotypes have been drawn to form the full first generation. This marks the frequency sample size $s$ as a hyperparameter. Note that the number of possible genotypes that can be drawn is given by a binomial coefficient, and due to the factorial in this coefficient this number will be very large. In order to explore a large section of the solution space, the sample size $s$ needs to be chosen proportional to both $g$ and the maximum runtime $r$ of the algorithm expressed in the number of generations. In the $k$-th generation, a maximum in the order of magnitude of $s \cdot 2^{k}$ frequencies are included in the genotype. Using this information an appropriate combination of values for $s$, $g$ and $r$ can be set.

\begin{table}[!]
\begin{tabular}{|cclc|}
    \hline
    \textbf{Parameter} & \quad \textbf{Type} & \quad \textbf{Description} & \quad \textbf{Value} \\ \hline
    $\delta$ & \quad Float & \quad Confidence & \quad $0$ \\ \hline
    $w$ & \quad Float & \quad Penalty weight & \quad $8$ \\ \hline
    $g$ & \quad Int & \quad Population size & \quad $6000$ \\ \hline
    $d$ & \quad Boolean & \quad Algorithm direction & \quad \texttt{True} \\ \hline
    $s$ & \quad Int & \quad Population sample size & \quad $6$ \\ \hline
    $r$ & \quad Int & \quad Number of generations & \quad $30$ \\ \hline
\end{tabular}
\caption{The collection of hyperparameters that defines the evolutionary algorithm. The value column contains the values for these parameters used for the realisation of the algorithm.}
\label{table:hyperparameters}
\end{table}

Recapitulating this subsection, besides the initial choices of background size during pre-processing and the probability measure for the parent selection function, the algorithm is defined by the hyperparameters $(\delta, w, g, d, s, r)$ given in Table \ref{table:hyperparameters}. The values in this table represent the hyperparameters used in this proof of concept, with \texttt{True} for the direction $d$ meaning the algorithm was ran backwards. This choice was made to prioritise accuracy, which also inspired the choice of $\delta$. The runtime $r$ was set to well above the observed limit of stagnation in trial runs, with the population size $g$ and frequency sample size $s$ empirically determined following the previous discussion. The penalty weight $w$ is set to $8$, but any value to counterbalance the increasing modulus in the objective function as the evolutionary algorithm progresses and frequencies are removed would have sufficed. These choices of hyperparameters consistently drive the evolutionary algorithm to converge to similar and well-performing frequency array phenotypes on repeated runs, suggesting the optimisation does not get stuck in easily avoidable local minima. The runtime for the algorithm (including pre-processing) on an Intel Gold 6154 CPU is in the order of magnitudes of minutes, \qm{bound by the longest recorded run.}

\subsection{Performance measures}\label{subsec:performancemeasures}

In the remainder of this section two metrics to measure performance will be introduced. The first is a fractional measure of the speedup obtained by using a frequency array phenotype output by the evolutionary algorithm. The second measures the loss in accuracy for a frequency array, where the exact waveform was injected into LIGO Livingston detector noise before the SNR was computed by matched filtering using both the exact and approximate waveforms as templates. The percentual difference in SNR is then used as a measure of loss.

For the definition of the speedup metric, let $t_{a}$ be the average time in seconds elapsed by the approximant per frequency, and $t_{i}$ the average time required to interpolate at a frequency in seconds. With $\rho := \# F / \# \mathbb{F}$ the relative size of the frequency array, the speedup of $F$ can be computed as:

\begin{equation}
    S[F] = \frac{\#\mathbb{F} \cdot t_{a}}{\#\mathbb{F} \cdot \left[ \rho t_{a} + (1 - \rho) t_{i} \right]},
\end{equation}

\noindent since the numerator specifies the time if the waveform is computed at every frequency, and the denominator specifies the time if the waveform is computed at a fraction $\rho$ of the frequencies, with interpolation at a fraction $(1 - \rho)$ of the frequencies. This expression reduces to:

\begin{equation}
    S[F] = \frac{1}{(1 - \rho) \cdot t_{i} / t_{a} + \rho}.
    \label{eq:speedup}
\end{equation}

\noindent Note that this quantity is dimensionless, measuring the speedup as a ratio in an isolated section of a procedure where the waveform is generated, without depending on the machine. In order to obtain accurate values for $t_{a}$ and $t_{i}$, $100$ waveforms were generated on the full spectrum using \texttt{Bilby} \cite{Ashton_2019}, and the same $100$ waveforms were interpolated on the frequency array phenotype. For every frequency array the parameters of the corresponding reference event were used without loss of generality since only time is measured, with the primary mass perturbed by $\pm 1$ at every iteration to ensure no values were cached. The times per frequency were then averaged out for each event and type, repeating and averaging this process $100$ times to remove possible inconsistencies.

Finding the neighbourhood in the parameter space $\mathcal{P}$ where the frequency arrays are effective is important, as the reference event input to the algorithm during pre-processing adds a level of specificity. Nearby the reference events the frequency arrays are likely to allow a more accurate reconstruction of the SNR. In order to analyse the generalisibility, $10,000$ points are sampled from the default precessing binary black hole prior in \texttt{Bilby} version 2.2.2. At every sampled set of parameters, the exact plus-polarisation of the waveform is generated and injected into simulated noise generated using the LIGO Livingston PSD included in Bilby version 2.2.2. The SNR of both the exact plus-polarisation, hereafter also called the exact SNR, and the approximate plus-polarisation, hereafter called the approximate SNR or $\textup{SNR}^{*}$, is computed against this strain. The percentual error for the frequency array $F$ is then calculated as:

\begin{equation}
    E_{\%}[F] = \frac{\left|\textup{SNR} - \textup{SNR}^{*}\right|}{\textup{SNR}} \cdot 100.
    \label{eq:percerror}
\end{equation}

\noindent This metric is computed as a percentage to account for the different values of the exact SNR. A given absolute error might be negligible for large SNR values, whereas it might be detrimental to a low value SNR reconstruction. Computing the error as a percentage removes this consideration. A threshold of $0.01$ is set for $E_{\%}[F]$, and a parameter sample is said to be rejected if the percentual error either attains or exceeds this threshold. Considering the uncertainties in SNR values in the GWTC-3 catalog \cite{theligoscientificcollaboration2021gwtc3} this threshold is particularly strict. This treshold is used as an aid to quantise results and is inconsequential to their objective interpretation.

%%%%---------------------------------------------------
\section{Results}\label{sec:results}

In this section the results from running the evolutionary algorithm with the three reference events GW150914, GW190412 and GW190814 are presented. The three frequency arrays output by the evolutionary algorithm are analysed in terms of speedup, accuracy and their generalisibility before selecting one of the three based on how well these three qualities are balanced. This frequency array is used for additional tests.

\begin{figure}[!]
    \includegraphics[width=1\columnwidth]{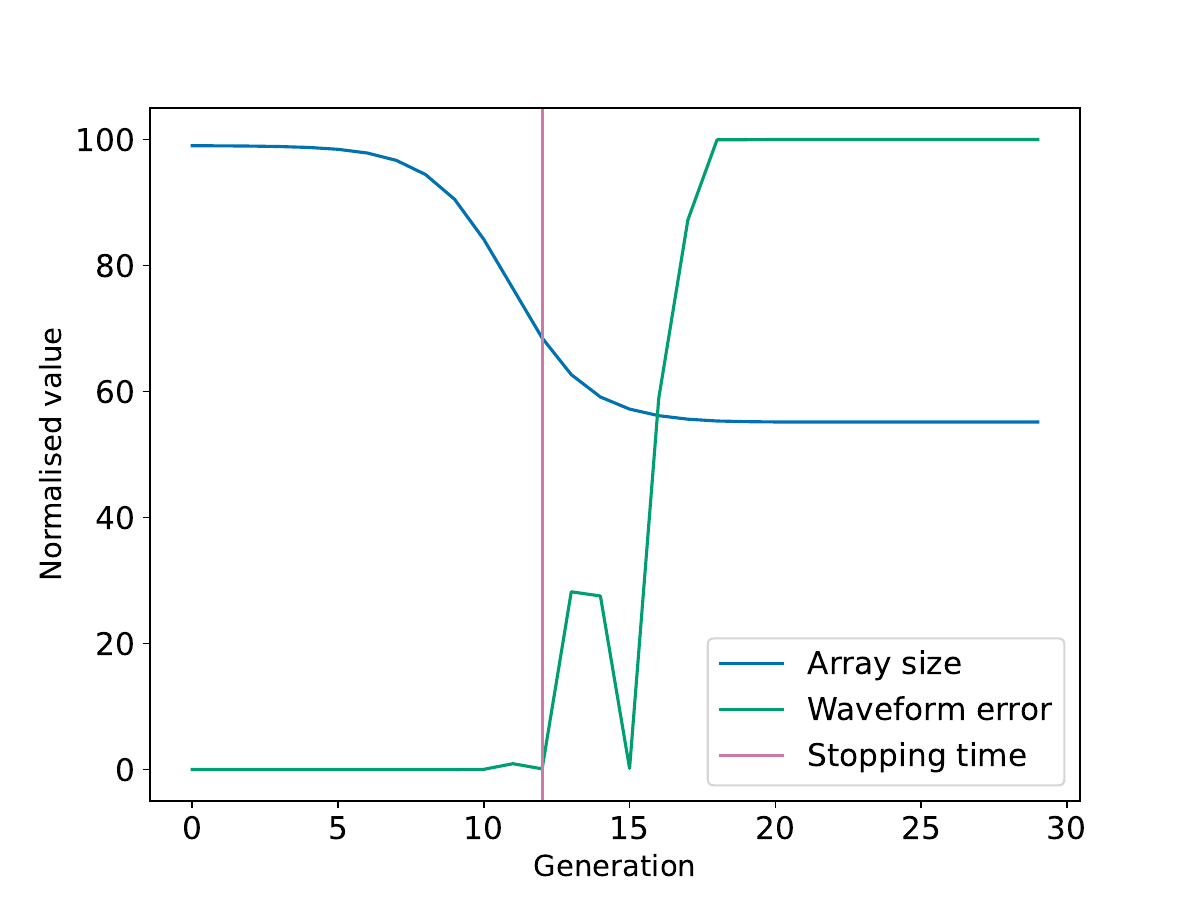}%
    \caption{A visualisation of the two parts that make up the objective function for GW150914, along with the stopping time for this event (generation 12). The waveform error is normalised to $[0, 100]$ so that the graphs can be compared at the same scale.}
    \label{fig:GW150914_tradeoff}
\end{figure}

\begin{figure}[!]
    \includegraphics[width=1\columnwidth]{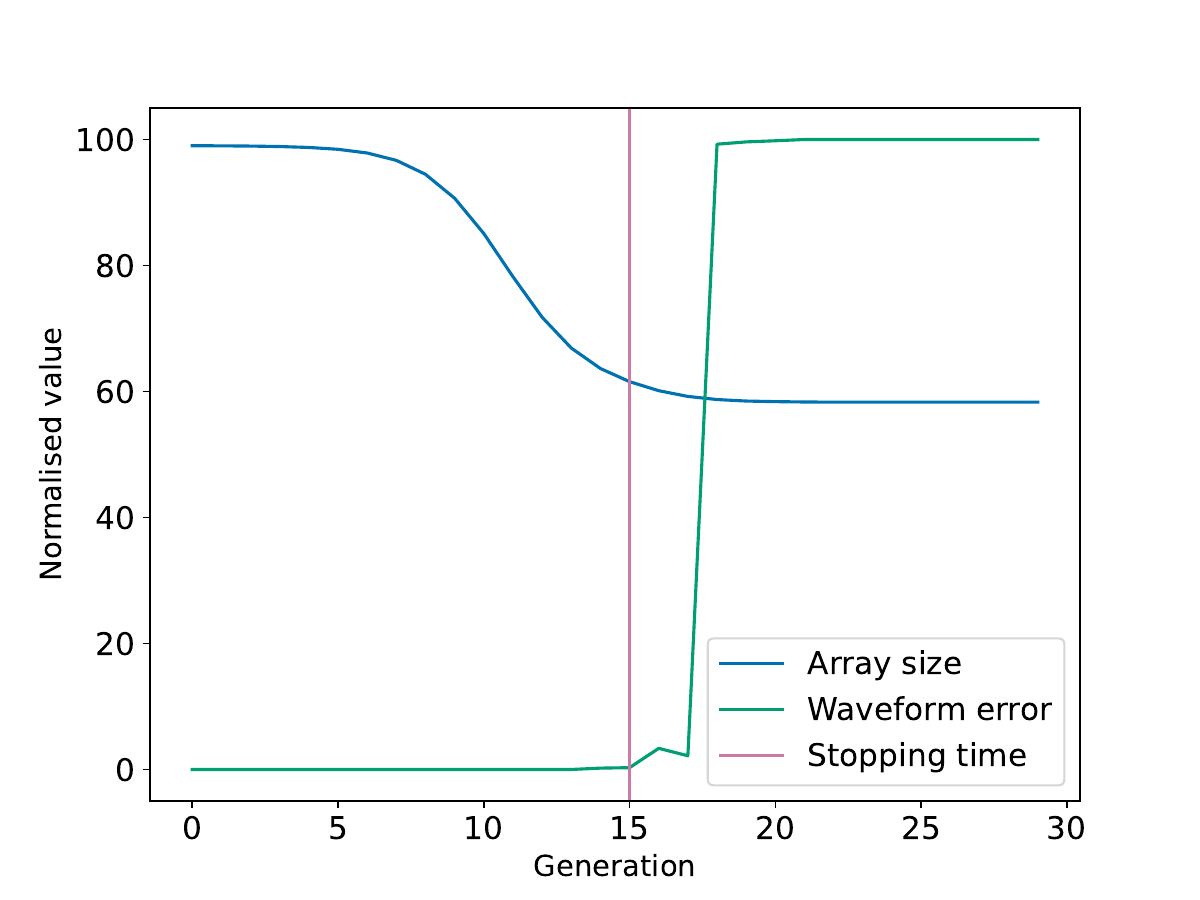}%
    \caption{A visualisation of the two parts that make up the objective function for GW190412, along with the stopping time for this event (generation 15). The waveform error is normalised to $[0, 100]$ so that the graphs can be compared at the same scale.}
    \label{fig:GW190412_tradeoff}
\end{figure}

\begin{figure}[!]
    \includegraphics[width=1\columnwidth]{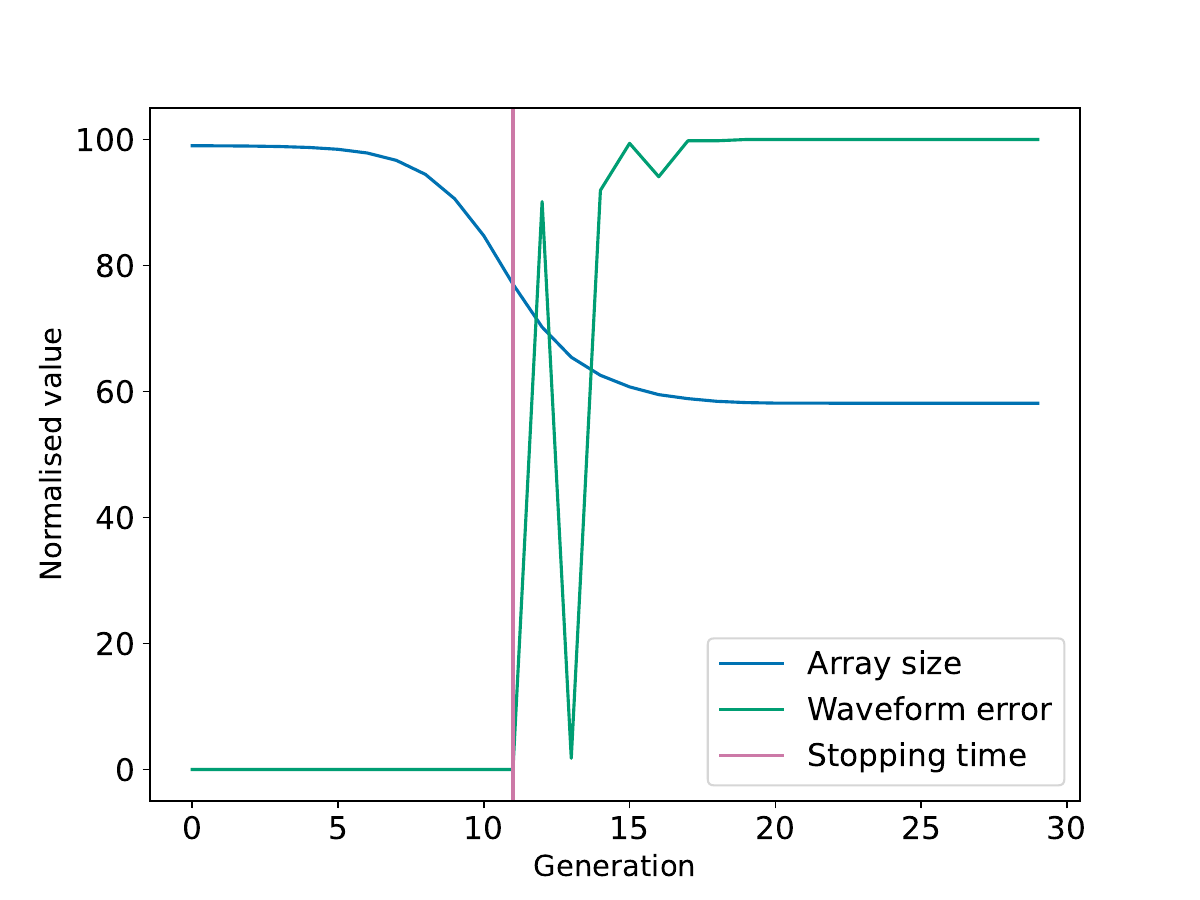}%
    \caption{A visualisation of the two parts that make up the objective function for GW190814, along with the stopping time for this event (generation 11). The waveform error is normalised to $[0, 100]$ so that the graphs can be compared at the same scale.}
    \label{fig:GW190814_tradeoff}
\end{figure}

The results for the reference events are shown in Fig. \ref{fig:GW150914_tradeoff}, \ref{fig:GW190412_tradeoff} and \ref{fig:GW190814_tradeoff} respectively, where both constituents of the objective function are shown in a visually separable way. The waveform error in these figures refers to the residual sum of squares, normalised to $[0, 100]$ so it is more easily compared to the percentual array size at this same scale. Normalisation here means the minimum achieved waveform error is mapped to $0$, and the maximum achieved waveform error is mapped to $100$. This means that the visualised waveform errors give little qualitative information and can not be compared across figures, because the original scales are not shown. The benefit however is that this allows for a visual rule to select the stopping time, and one that is in fact automatable: halting at the first sign of a significant jump in error. At this time the evolutionary algorithm is stopped and the frequency array with the highest fitness score is chosen as the solution.

%\begin{figure}[!]
%    \includegraphics[width=1\columnwidth]{GW190412_gen_29_reconstruction_real.pdf}%
%    \caption{The real part of the exact plus-polarised component of GW190412 waveform, with the difference to the approximate component on the frequency array proposed in the last generation ($29$) of the algorithm visualised. The horizontal axis is in log scale. \nh{It is better to denote the y-axis label as the amplitude} \comm{the residual plotted in orange will need to be zoomed in. I suggest including a small zoomed in panel under the waveform plot for just the orange line. The y-scale of this panel will probably need to be a few orders of magnitude smaller than the waveform panel. This same comment applies to Fig 5. Also, don't forget to include units on all of your plots and in your tables.}}
%    \label{fig:GW190412_real_reconstruction}
%\end{figure}

%\begin{figure}[!]
%    \includegraphics[width=1\columnwidth]{GW190412_gen_29_reconstruction_imag.pdf}%
%    \caption{The imaginary part of the exact plus-polarised component of GW190412, with the difference to the approximate component on the frequency array proposed in the last generation ($29$) of the algorithm visualised. The horizontal axis is in log scale.}
%    \label{fig:GW190412_imag_reconstruction}
%\end{figure}

\begin{figure*}[!]
    \includegraphics[width=1\textwidth]{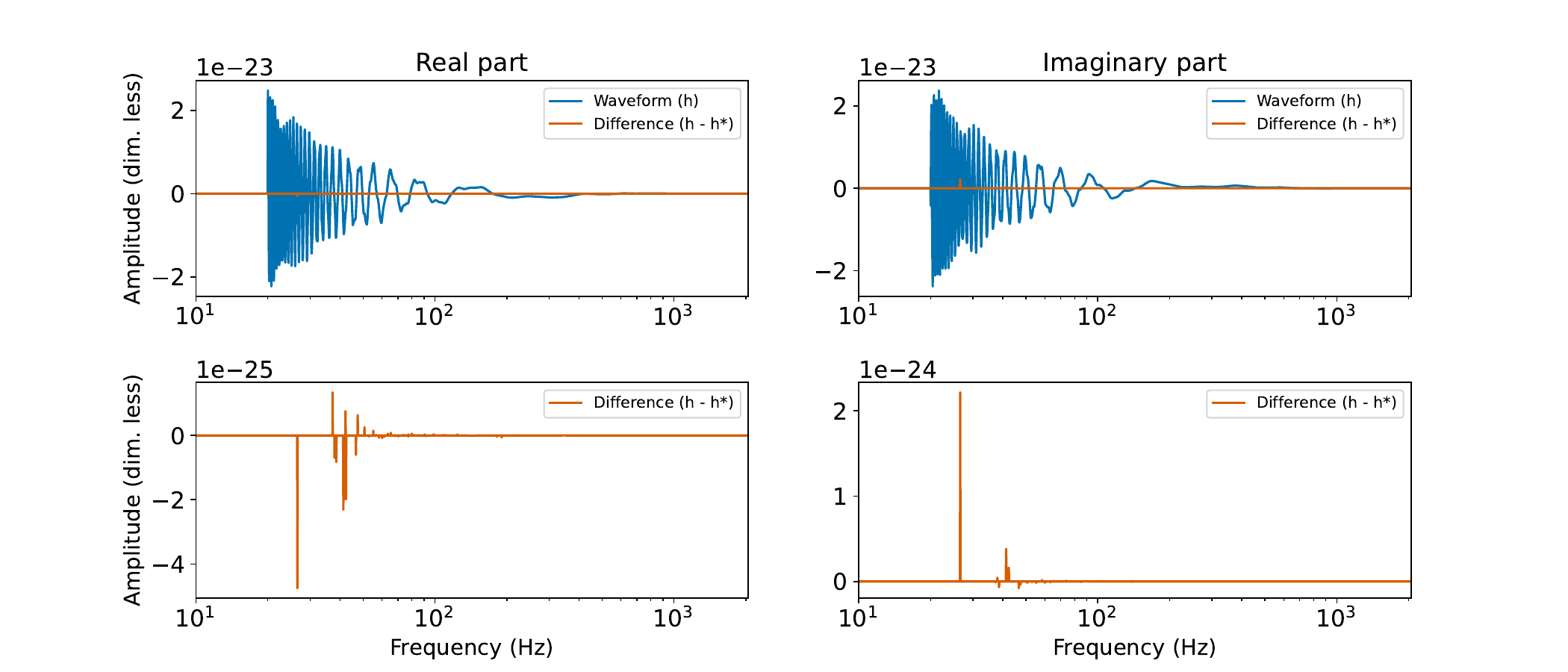}%
    \caption{The real and imaginary parts of the plus-polarised component of GW190412 ($h$) and the difference with the approximation ($h - h^{*}$) computed on the frequency array proposed in the last generation ($29$) of the algorithm. The bottom two panels show the differences on a smaller scale, and the horizontal axes are in log scale.}
    \label{fig:GW190412_reconstruction}
\end{figure*}

At first glance, the non-monotonicity of the waveform error might seem conflicting with the assumption that every subsequent generation contains better adapted solutions. However, one should keep in mind that this statement is made with respect to the objective function, where the waveform error is combined with a size penalty, and that the objective function will show monotonicity. Overall, the fact that the waveform error is only relative is underlined by the difference between the exact and approximate waveform visualised in Fig. \ref{fig:GW190412_reconstruction}. Despite the dramatic jump in waveform error seen in Fig. \ref{fig:GW190412_tradeoff}, the difference $h-h^{*}$ for the frequency array proposed in the last generation (where the error is maximised) barely registers at the scale of the waveform, being at least an order of magnitude smaller. These magnitudes can be considered as upper bounds for the waveform errors over the runtime of the evolutionary algorithm.

\qm{In the current implementation, the error is more pronounced in the phase than it is in amplitude. Depending on the application, the objective function can be modified to either balance this error, or favor one of the two.}

\begin{figure}[!]
    \includegraphics[width=1\columnwidth]{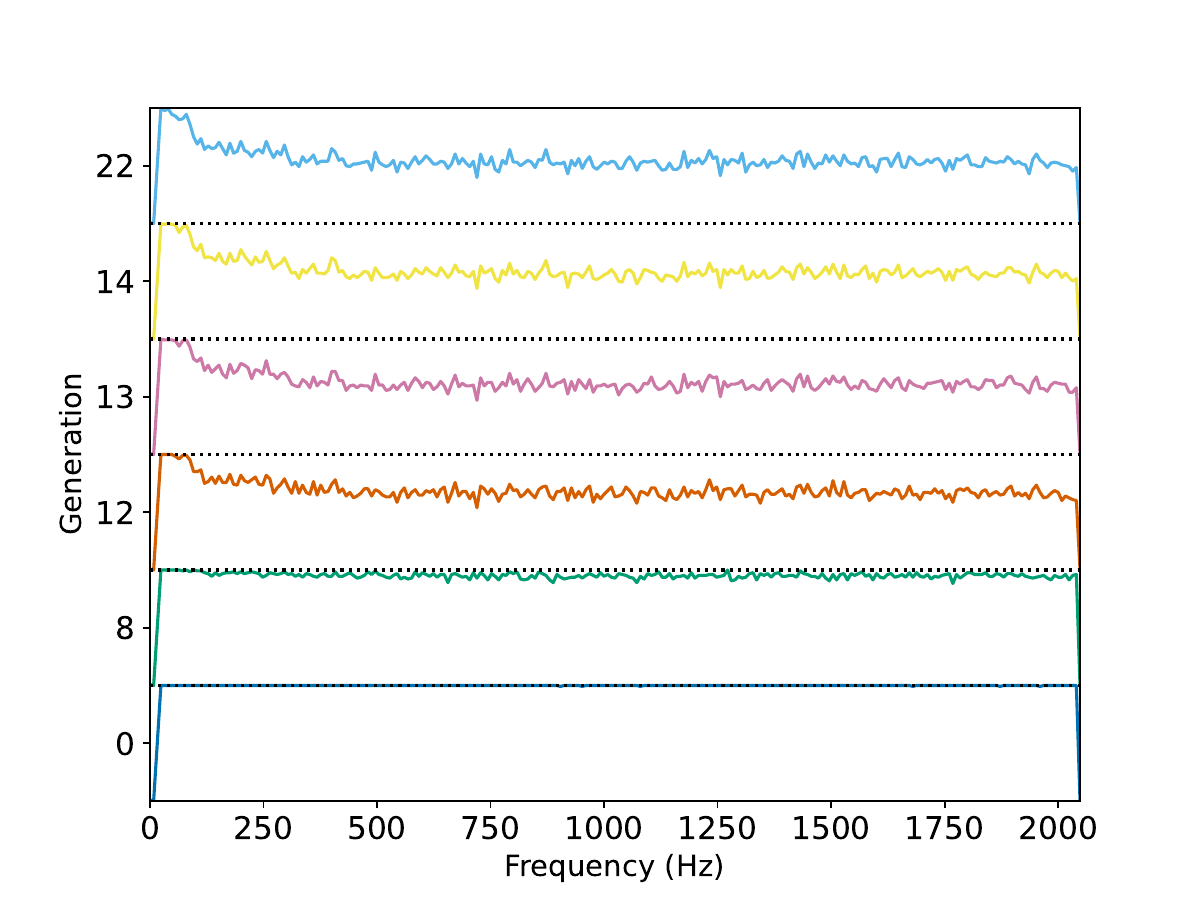}%
    \caption{The evolution of the proposed frequency arrays for GW150914 visualised for a selection of generations. The densities are computed by binning the frequency spectrum, counting frequencies, and normalising to the maximum count. The evolution can be seen to follow the trend in the array size seen in Fig. \ref{fig:GW150914_tradeoff}, and shows what frequencies are removed during the runtime of the evolutionary algorithm. Notably, the frequencies between $0$ and $250$ Hz persist the longest.}
    \label{fig:GW150914_densities}
\end{figure}

The typical development of the proposed frequency array for GW150914 over the generations is visualised in Fig. \ref{fig:GW150914_densities}. The generations shown were selected to equally space the process before stagnation occurs at generation $22$, and includes the ultimately selected candidate from generation $12$. Both the stagnation and jump in waveform error that underlies this choice of candidate can be seen in Fig. \ref{fig:GW150914_tradeoff}. At every generation, a distribution function over the frequency spectrum is shown. This distribution is computed by dividing the spectrum into bins, counting the frequencies present in each bin, and normalising the counts to the maximum count over all bins. Mirroring the development of the array size in Fig. \ref{fig:GW150914_tradeoff}, few frequencies are removed from the early candidates, with only a minor visible move downwards from the dotted line seen in generation $8$. Change accelerates between generations $8$ and $12$. Note that from the peaks and valleys that persist from this generation onwards, the algorithm has decided at an early stage which frequency ranges are most important. The lower end of the frequencies are kept consistently, although the dropoff towards $250$ Hz becomes steeper during the later generations. Interpreting this figure gives an indication of how the evolutionary algorithm has traversed the search space, favoring the lower range of the spectrum, but always locally to the extent that above $250$ Hz, the densities look like a diffused horizontal line. This means the evolutionary algorithm does not favor any band of the spectrum as strongly as it does $[0, 250]$ Hz.

\begin{table}[!]
\begin{tabular}{|lcccc|}
    \hline
    \textbf{} & \quad \textbf{$\rho$} & \quad \textbf{$t_{a}$} & \quad \textbf{$t_{i}$} & \quad \textbf{S} \\ \hline
    \textbf{GW150914} & \quad 0.69 & \quad 5.52e-07 & \quad 6.95e-12 & \quad 1.46 \\ \hline
    \textbf{GW190412} & \quad 0.62 & \quad 8.48e-07 & \quad 7.66e-12 & \quad 1.62 \\ \hline
    \textbf{GW190814} & \quad 0.77 & \quad 1.71e-06 & \quad 5.90e-12 & \quad 1.30 \\ \hline
\end{tabular}
\caption{Measurements of the values required for the computation of the speedup $S[F]$ given in Eq. \ref{eq:speedup}, along with the computed speedup. These values are $\rho$, the size of the frequency array relative to the spectrum, $t_{a}$, the average time in seconds spent computing the waveform at a frequency by the approximant and $t_{i}$, the average time in seconds spent to interpolate the waveform at a frequency. All values are rounded to two decimals for readability.}
\label{table:speedup}
\end{table}

The speed-up of the frequency arrays are shown in Table \ref{table:speedup}, showing the input variables and values for $S[F]$ as defined in Eq. \ref{eq:speedup}. A detail that should be noted is that there is a natural inverse relation between the relative array size $\rho$ and speedup $S$. The more frequencies are omitted from the frequency array, the higher the speedup. However, the values in the time columns differ to an extent. The differences in the $t_{i}$ column can be explained by the density of the frequencies included in the array. The time measurements given in this column correspond to the average time elapsed interpolating at a frequency that is not included in the array. If many subsequent frequencies are omitted, the value at these frequencies can be interpolated at once from a single linear segment, whereas the same number of values for non-subsequent frequencies need to be interpolated using several such segments. The computation times scales linearly with the number of segments that are computed and will therefore be higher in the latter scenario. For the $t_{a}$ column, the time difference between frequency arrays is likely to stem from the additional effects added to the waveforms, such as twisting up. 

\qm{As a note on data compression and memory usage, before interpolating to the full frequency spectrum, one can opt to write the waveform to disk, indexed by the frequency array. By choosing to interpolate to the full spectrum later, waveforms can be stored in a compressed format, and the compression factor will equal $\rho$.}

\begin{table*}[!]
\begin{tabular}{|lcccc|}
    \hline
    \textbf{} & \quad \textbf{Range} & \quad \textbf{GW150914} & \quad \textbf{GW190412} & \quad \textbf{GW190814} \\ \hline
    \textbf{Mass ratio} & \quad $[0.125, 1]$ & \quad $[0.125, 1]$ & \quad $[0.125, 1]$ & \quad $[0.126, 1]$ \\ \hline
    \textbf{Chirp mass ($M_{\odot}$)} & \quad $[25, 100]$ & \quad $[25.005, 86.922]$ & \quad $[25.007, 86.855]$ & \quad $[25, 86.509]$ \\ \hline
    \textbf{Luminosity distance (Mpc)} & \quad $[100, 5000]$ & \quad $[101.091, 4999.683]$ & \quad $[170.862, 4999.658]$ & \quad $[128.847, 4999.707]$ \\ \hline
    \textbf{Declination (Rad)} & \quad $[-1.571, 1.571]$ & \quad $[-1.552, 1.557]$ & \quad $[-1.553, 1.553]$ & \quad $[-1.545, 1.559]$ \\ \hline
    \textbf{Right ascension (Rad)} & \quad $[0, 6.283]$ & \quad $[0, 6.282]$ & \quad $[0, 6.281]$ & \quad $[0, 6.283]$ \\ \hline
    \textbf{\qm{Orbital inclination} (Rad)} & \quad $[0, 3.142]$ & \quad $[0.003, 3.114]$ & \quad $[0.005, 3.138]$ & \quad $[0.026, 3.121]$ \\ \hline
    \textbf{Polarisation angle (Rad)} & \quad $[0, 3.142]$ & \quad $[0, 3.141]$ & \quad $[0, 3.141]$ & \quad $[0.001, 3.141]$ \\ \hline
    \textbf{Coalescence phase (Rad)} & \quad $[0, 6.283]$ & \quad $[0.001, 6.283]$ & \quad $[0, 6.283]$ & \quad $[0, 6.282]$ \\ \hline
    \textbf{Spin of primary mass} & \quad $[0, 0.99]$ & \quad $[0, 0.99]$ & \quad $[0, 0.99]$ & \quad $[0, 0.99]$ \\ \hline
    \textbf{Spin of secondary mass} & \quad $[0, 0.99]$ & \quad $[0, 0.99]$ & \quad $[0, 0.99]$ & \quad $[0, 0.99]$ \\ \hline
    \textbf{Tilt of primary mass (Rad)} & \quad $[0, 3.142]$ & \quad $[0.014, 3.136]$ & \quad $[0.023, 3.126]$ & \quad $[0.018, 3.1]$ \\ \hline
    \textbf{Tilt of secondary mass (Rad)} & \quad $[0, 3.142]$ & \quad $[0.004, 3.13]$ & \quad $[0.026, 3.127]$ & \quad $[0.017, 3.125]$ \\ \hline
    \textbf{Precession angle (Rad)} & \quad $[0, 6.283]$ & \quad $[0, 6.283]$ & \quad $[0.003, 6.283]$ & \quad $[0, 6.283]$ \\ \hline
    \textbf{Azimuthal angle (Rad)} & \quad $[0, 6.283]$ & \quad $[0, 6.282]$ & \quad $[0, 6.282]$ & \quad $[0, 6.283]$ \\ \hline
\end{tabular}
\caption{The ranges for each of the variables in the parameter space and the recovered ranges per frequency array, \qm{in rounded numerical format}. If no unit is included, the quantity is dimensionless. Note that the geocent time is not included, as it is inconsequential to this test. All parameter ranges save for the luminosity distance and chirp mass were recovered, with both of these not being fully covered by the drawn sample of parameters.}
\label{table:samples}
\end{table*}

The results of the generalisibility tests to determine the regions of the parameter space $\mathcal{P}$ where the frequency arrays are effective are shown in Table \ref{table:samples}. For each of the $15$ parameters (minus the geocent time) the sampled ranges are shown, as well as the ranges that were recovered for the frequency arrays. Recovered here means that the parameter samples from these ranges were accepted by their error being below the given threshold, with $E_{\%}[F] < 0.01$. For most parameters, the full ranges were recovered, save for the luminosity distance and chirp mass. In the case of the luminosity distance, further inspection reveals that the lower ranges were simply not reached in the parameter sample drawn. The same holds for the chirp masses due to the way \texttt{Bilby} samples the chirp mass through uniform sampling in the component masses. \qm{If this is manually changed the results might vary slightly.}

\begin{table}[!]
\begin{tabular}{|lccc|}
    \hline
    \textbf{} & \quad \textbf{Minimum} & \quad \textbf{Maximum} & \quad \textbf{Rejected} \\ \hline
    \textbf{GW150914} & \quad 2.69e-10 & \quad 0.50 & \quad 909 \\ \hline
    \textbf{GW190412} & \quad 1.12e-07 & \quad 2.87 & \quad 902 \\ \hline
    \textbf{GW190814} & \quad 1.47e-09 & \quad 0.21 & \quad 203 \\ \hline
\end{tabular}
\caption{The extrema for the percentual errors $E_{\%}[F]$ on the parameter sample, along with the number of parameter samples rejected, as for these rejected samples $E_{\%}[F] \geq 0.01$.}
\label{table:sample_error_bounds}
\end{table}

\begin{figure}[!]
    \includegraphics[width=1\columnwidth]{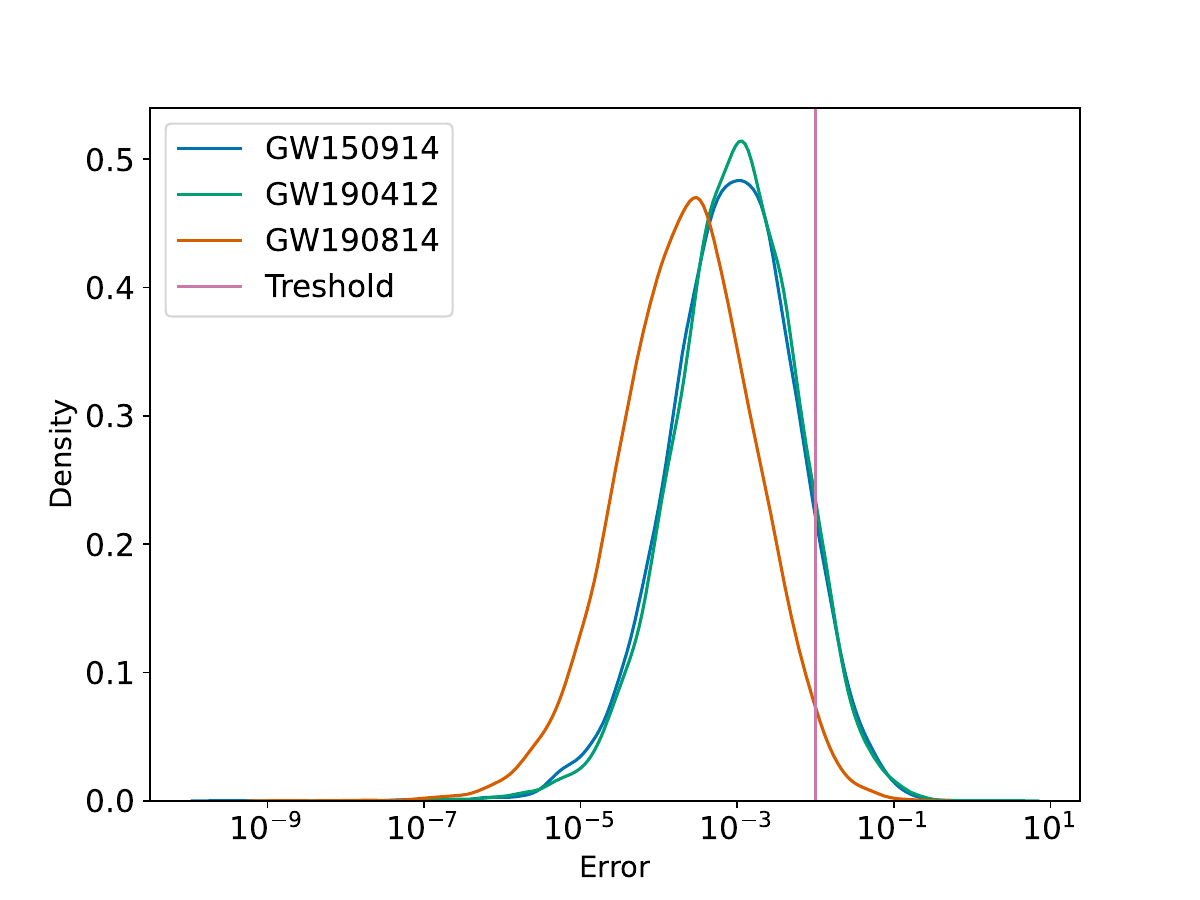}%
    \caption{Densities of the percentual error in SNR $E_{\%}[F]$ for the full parameter samples, along with the threshold for acceptance. The horizontal axis is in log scale.}
    \label{fig:sample_error_densities}
\end{figure}

Additional statistics on the generalisibility tests are shown in Table \ref{table:sample_error_bounds}. A number of parameter sample points were rejected. However, the errors for these samples were evaluated at a strict error threshold. Keeping in mind the generalisibility, these results can be considered remarkable, as the error is bounded above by $2.87\%$. The distribution of the errors is further visualised in Fig. \ref{fig:sample_error_densities}. This plot shows that for all three frequency arrays the percentual errors in SNR are mostly concentrated below the error threshold of $10^{-2} = 0.01$.

\begin{figure}[!]
    \includegraphics[width=1\columnwidth]{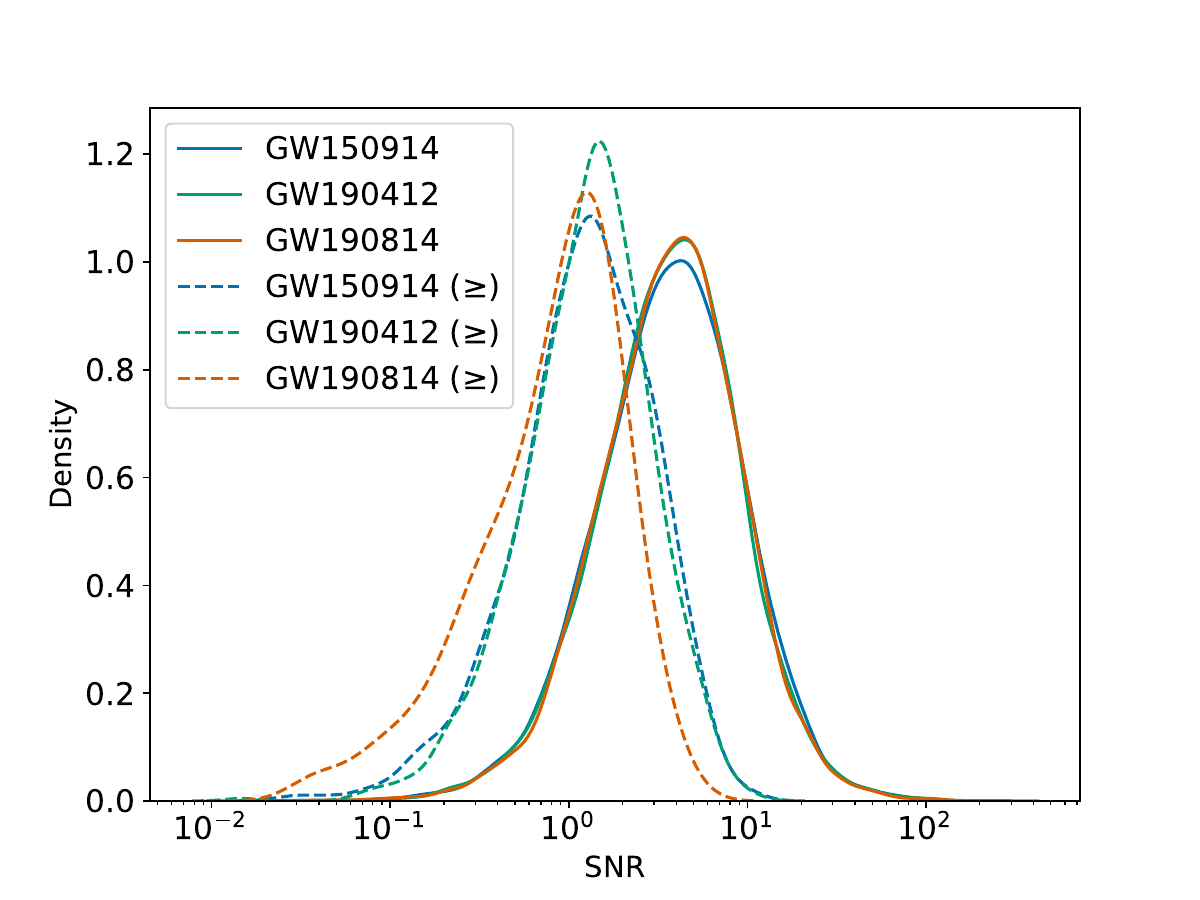}%
    \caption{The exact SNR distributions for the full parameter samples (solid) and the rejected parameter sample points that exceed $(\geq)$ the threshold of $0.01$ set on $E_{\%}[F]$ (dashed). The horizontal axis is in log scale.}
    \label{fig:sample_rejected_full_snr_distributions}
\end{figure}

Investigating the cause for the rejection of parameter samples, it becomes apparent that nearly all rejected samples have a low exact SNR value \qm{for $h$}. This is visualised in Fig. \ref{fig:sample_rejected_full_snr_distributions}, showing the exact SNR distributions for the rejected parameter samples versus the exact SNR distributions for the full samples. It can be seen from this figure that for the rejected parameter samples, the SNR is on average an order of magnitude smaller than that of the full parameter sample, and moreover that the distributions are left-tail heavy. In order to verify the strength of the SNR as a predictor for rejection, logistic regression \cite{10.5555/1162264} of the binary acceptance class onto the waveform parameters was employed. The sampled parameters for the three candidates were stacked, with the luminosity distance replaced by the SNR. The removal of the luminosity distance is a requirement so that the multicollinearity assumption of logistic regression is not violated, as luminosity distance correlates strongly with SNR. Despite the SNR scaling with several other parameters, the removal of the luminosity distance should sufficiently reduce the correlation. A binary labelling was created for this dataset by assigning a dependent variable $1$ to the accepted parameter samples and $0$ to the rejected samples. Fitting a logistic regression model, the SNR was found to be the predictor with the highest coefficient by far, with a value of $\beta_{\textup{SNR}} = 16.5$. This coefficient was followed by the chirp mass with a value of $\beta_{\textup{Chirp}} = 3.74$, meaning the SNR is a significantly stronger predictor for the rejection of a parameter sample according to the regression model. As with accuracy, the generalisibility too is bounded from below by the results presented.

%\begin{figure*}[!]
%    \includegraphics[width=1\textwidth]{GW150914_partial_sample_heatmap_tight.pdf}%
%    \caption{Heatmaps for the absolute errors in SNR of ten partial samples when using the proposed frequency array learned on GW150914. A partial sample means that a sample was drawn from the parameter space without sampling the chirp mass or mass ratio. For every such sample, the chirp mass and ratio were varied within the ranges of $[25, 100]$ and $[0.125, 1]$ respectively, and the errors are visualised in the heatmaps shown here.}
%    \label{fig:GW150914_partial_sample_heatmap}
%\end{figure*}

\begin{figure}[!]
    \includegraphics[width=1\columnwidth]{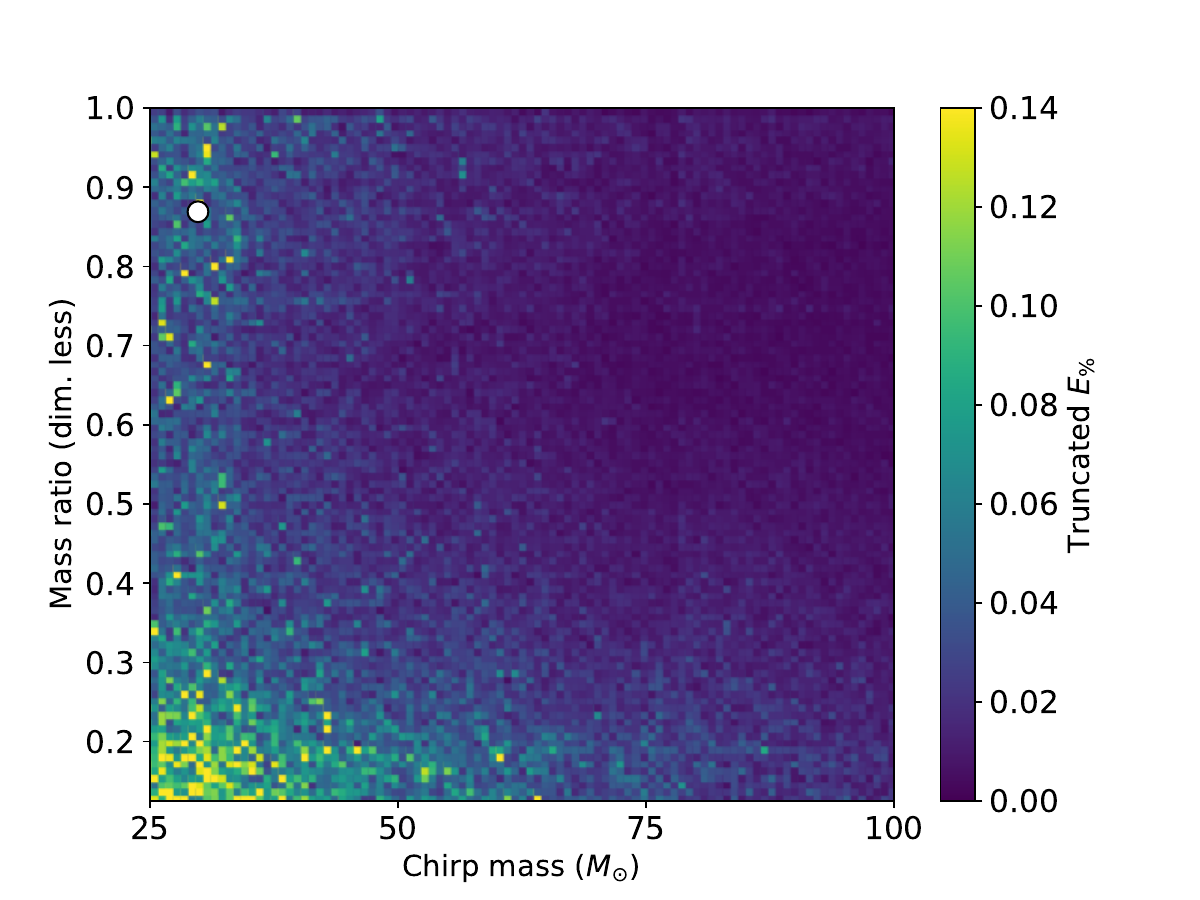}%
    \caption{A heatmap for the error in SNR $E_{\%}$ based on partial parameter samples when using the proposed frequency array obtained using GW150914 as the reference event. A partial parameter sample means that a sample was drawn from the parameter space without sampling the chirp mass or mass ratio. For every such sample, the chirp mass and mass ratio were varied within the ranges of $[25, 100]$ and $[0.125, 1]$ respectively. The average errors for ten such samples are shown in this figure, truncated at $0.14\%$ for increased visibility. The mass ratio and chirp mass of GW150914 are indicated by a white marker.}
    \label{fig:GW150914_partial_sample_heatmap_perc_average}
\end{figure}

Based on the analysis in this section, the frequency array obtained by using GW150914 as the reference event is proposed as a proof of concept, as it obtains a speed-up in computational time of $46\%$ at a maximum error of $0.5\%$ in SNR. A heatmap for the generalisibility of this frequency array is shown in Fig. \ref{fig:GW150914_partial_sample_heatmap_perc_average}. This figure was created by drawing $10$ partial parameter samples, meaning every parameter except for the mass ratio and chirp mass were sampled, and then varying the mass ratio and chirp mass before computing the \qm{percentual} error \qm{in SNR} $E_{\%}$ for every such combination. A point in the plane then corresponds to the average error of the $10$ parameter samples for that combination of the chirp mass and mass ratio, and the full figure shows $100,000$ such parameter samples in total. It can be seen from the figure that especially in the band of the chirp mass where the event belongs to, moving away from the mass ratio of the event will increase the error.

%\begin{figure}[!]
%    \includegraphics[width=1\columnwidth]{GW150914_partial_sample_heatmap_average.pdf}%
%    \caption{xyz}
%    \label{fig:GW150914_partial_sample_heatmap_average}
%\end{figure}

\begin{figure}[!]
    \includegraphics[width=1\columnwidth]{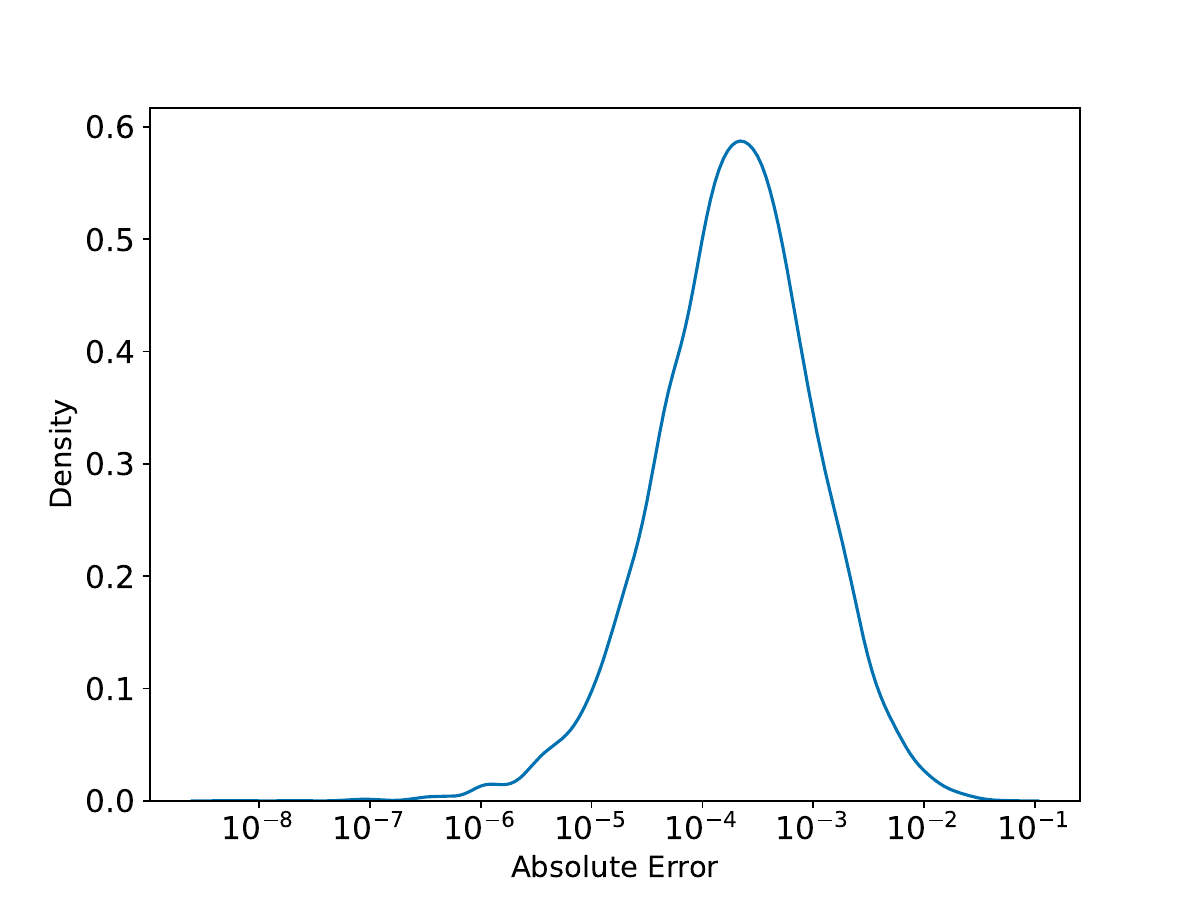}%
    \caption{The distribution of the absolute error $\left| \mathcal{L}_{m} - \mathcal{L}^{*}_{m} \right|$ between the exact log likelihood maximum $\mathcal{L}_{m}$ computed using the exact waveform and the approximate log likelihood maximum $\mathcal{L}^{*}_{m}$ computed using the approximate waveform. This distribution was constructed on a sample of size $10,000$. The horizontal axis is in log scale.}
    \label{fig:GW150914_likelihood_error_distribution}
\end{figure}

As an additional test for the applicability of the proof of concept, the maxima of the log likelihoods \cite{PhysRevD.81.062003} that quantify the likelihood of waveforms being present in detector strains were approximated using the frequency array. Using the inner product defined in Eq. \ref{eq:innerproduct}, the log likelihood can be defined as:

\begin{equation}
    \mathcal{L}(h, D) = - \frac{1}{2} \langle D - h, D - h \rangle
\end{equation}

\noindent for $h$ the waveform projected onto the detector frame and $D$ the strain. Setting $D := N + h'$ for $N$ noise, the maximum of the log likelihood should be attained at the injected waveform $h'$. Analogous to the SNR reconstruction, the exact log likelihood maximum $\mathcal{L}(h, N+h)$, or $\mathcal{L}_{m}$ for shorthand, and approximate log likelihood maximum $\mathcal{L}(h^{*}, N+h)$, or $\mathcal{L}^{*}_{m}$, were computed for a parameter sample of size $10,000$. Here, however, both polarisations are included. Note that these errors do not need to be converted to percentages as was done for the errors in SNR, since the variance in the values will be significantly smaller. The absolute error distribution is shown in Fig. \ref{fig:GW150914_likelihood_error_distribution} and shows promising results. The reader should be aware however that the evolutionary algorithm was not finetuned to this computation, suggesting that a higher accuracy can likely be obtained. Specifically, the objective function \qm{(that defines the optimisation problem)} should be adapted to the log likelihood function, and the hyperparameters should be rebalanced.

Further study in general might reveal marginally better performing frequency arrays, or frequency arrays better suited to specific sections of the parameter space. It is however expected that the improvement from frequency subsampling and linear interpolation represented by the values in this section is near-optimal. Implementing the naive frequency subsampling scheme of interpolating every second frequency (setting \qm{the relative array size at} $\rho = 0.5$) for GW150914, the speedup $S$ evaluates to $1.99$, but the minimum error of $E_{\%}$ blows up by a factor with order of magnitude $10^{4}$, in the process causing a rejection for 5972 parameter samples.

It follows from the results in this section that as expected, for a given threshold, smaller frequency arrays are effective on smaller neighbourhoods of the parameter space $\mathcal{P}$. Under the assumption of continuity, parameter instances nearby the reference event parameters will give better performance because of their similarity. If evolutionary algorithms are to be used to speed up computations on $\mathcal{P}$ with a given maximum allowable error, a single frequency array might not suffice. One option would then be to assign different frequency arrays to different regions of $\mathcal{P}$. In order to do this efficiently, the effective neighbourhood of an array will have to be computed. A first proposition could be akin to the generalisability test used in this paper. After setting a treshold, recover the ranges in which all parameter samples were accepted, and take the convex hull of these ranges in $\mathcal{P}$. There exist efficient algorithms for the computation of convex hulls in higher dimensions under the assumption of flat space \cite{preparata2012computational} that enable this procedure. Finding an optimal coverage of $\mathcal{P}$ by the resulting polytopes, however, reduces to an instance of the sphere packing problem through a topological argument, which depending on specifics is known to be NP-complete \cite{FOWLER1981133}. This obstacle can be overcome at the cost of efficiency if overlap of the polytopes is allowed and a global optimum is not required. Furthermore, for sufficiently large polytopes, the largest embeddable hypercube can be determined and used as an approximation to the polytope. These cubes in turn can be used to tile $\mathcal{P}$ more easily due to their shape. If such a tiling of $\mathcal{P}$ is generated the speedups demonstrated in this section can serve as lower bounds for the speedups achievable on the entirety of $\mathcal{P}$.

%%%%---------------------------------------------------
\section{Conclusions}\label{sec:conclusions}

Evolutionary algorithms were introduced as a method for the optimisation problem of subsampling frequency spectra to speed-up waveform approximants, and per extension, the computation of SNR. It was shown that even with strict bounds on the percentual errors, speedups of at least $30\%$ in computational time are achievable on nearly the entire sampled parameter ranges. If the error bound is heightened to at most $2.87\%$ the frequency arrays generalise to the full parameter ranges (specified in Table \ref{table:samples}). The margins of the parameter ranges where the frequency arrays were not effective were explored, as well as the possibility of tiling the parameter space to force a set upper bound on the error. It was demonstrated that the ranges where the frequency arrays were ineffective are mostly determined through correlation with the SNR, and it could therefore be argued that the frequency arrays generalise to the full ranges depending on the error metric in use. The proposed proof of concept frequency array, obtained by using GW150914 as a reference event, shows a speedup of $46\%$ with a maximum error of $0.5\%$ on the full parameter ranges. Additionally, it was shown that this frequency array performs exceptionally well for the reconstruction of the maxima of log likelihood functions.

Possible future lines of work include improving the methods presented in this paper specifically for parameter estimation \cite{Thrane_Talbot_2019} or time domain problems such as the fast computation of Euler angles that describe the evolution of precessing binary black hole systems \cite{PhysRevD.103.104056, PhysRevD.86.104063, PhysRevD.91.024043, PhysRevD.106.024020}. Alternatively, the tiling of the parameter space, and simultaneously the optimum in trade-off between accuracy and generalisibility, can be studied further. The evolutionary algorithm can in theory also be calibrated to perform on targeted sections of the parameter space, such as those corresponding to intermediate-mass black hole binaries or extreme mass ratio inspirals. Within these sections a single frequency array might suffice.

Further improvements can be made to the evolutionary algorithm, for instance by incorporating more prior information. The algorithm can be discouraged from traversing sections of the solution space, for example by using a supplied power spectral density, or by more hyperparameter tuning. Currently the algorithm is hardcoded to run for a total of $30$ generations. Despite a stopping condition being in place, the algorithm was ran for the full duration to study its behaviour. By tuning the hyperparameters and designing more advanced stopping conditions, the evolutionary algorithm can be modified to propose frequency arrays meeting a set of required conditions without additional user input. A future milestone is to release the adjustable code as a package.

\qm{For the evolutionary algorithm itself, the investigation of its applications to data compression for gravitational-wave data analysis could prove useful. This can be coupled to the exploration of alternative interpolation schemes.}

With the number of gravitational-wave detections increasing through the arrival of detector upgrades and new generations of observatories, the need for faster yet accurate analyses is becoming more apparent. The use of optimisation schemes such as evolutionary algorithms to realise these targets are indispensable, and will enable the scientific community to learn more about the nature of gravitational waves in less time. 

%%%%---------------------------------------------------
\section*{Acknowledgements}

With thanks to Harsh Narola, Stefano Schmidt, Melissa Lopez, Justin Janquart, Bhooshan Gadre, Marc van der Sluys, Chris van den Broeck, Justin Perez \qm{and the anonymous referee}. Q.M is supported by the research program of the Netherlands Organisation for Scientific Research (NWO). S.C is supported by the National Science Foundation under Grant No. PHY-2309332. The authors are grateful for computational resources provided by the LIGO Laboratory and supported by the National Science Foundation Grants No. PHY-0757058 and No. PHY-0823459. This material is based upon work supported by NSF's LIGO Laboratory which is a major facility fully funded by the National Science Foundation.

%%%%---------------------------------------------------
%\appendix
%\section{Partial Samples}

%%%%---------------------------------------------------

\bibliography{references}

\onecolumngrid

\end{document}